\documentclass{article}
\usepackage{spconf,amsmath,graphicx,hyperref,typeface,float,enumitem,multirow,booktabs,array,amssymb,tabularx}

\newcolumntype{L}[1]{>{\raggedright\arraybackslash}p{#1}}
\newcolumntype{C}[1]{>{\centering\arraybackslash}p{#1}}
\newcolumntype{R}[1]{>{\raggedleft\arraybackslash}p{#1}}

\title{IMPROVING ADVERSARIAL WAVEFORM GENERATION BASED SINGING VOICE CONVERSION WITH HARMONIC SIGNALS}
\name{Haohan Guo$^*$, Zhiping Zhou$^\dag$$^\ddag$, Fanbo Meng$^\dag$$^\ddag$, Kai Liu$^\dag$$^\ddag$}
\address{$^*$The Chinese University of Hong Kong, Hong Kong SAR, China \\
  $^\dag$Sogou Inc., Beijing, China, $^\ddag$Tencent PCG AIBU\\
  \ninept\href{mailto:hguo@se.cuhk.edu.hk}{\nolinkurl{hguo@se.cuhk.edu.hk}},
  \href{mailto:kevinllliu@tencent.com}{{\nolinkurl{{merryzhou,fanbomeng,kevinllliu}@tencent.com}}}
}

\begin{document}
%\ninept
%
\maketitle
\begin{abstract}
Adversarial waveform generation has been a popular approach as the backend of singing voice conversion (SVC) to generate high-quality singing audio. However, the instability of GAN also leads to other problems, such as pitch jitters and U/V errors. It affects the smoothness and continuity of harmonics, hence degrades the conversion quality seriously. This paper proposes to feed harmonic signals to the SVC model in advance to enhance audio generation. We extract the sine excitation from the pitch, and filter it with a linear time-varying (LTV) filter estimated by a neural network. Both these two harmonic signals are adopted as the inputs to generate the singing waveform. In our experiments, two mainstream models, MelGAN and ParallelWaveGAN, are investigated to validate the effectiveness of the proposed approach. We conduct a MOS test on clean and noisy test sets. The result shows that both signals significantly improve SVC in fidelity and timbre similarity. Besides, the case analysis further validates that this method enhances the smoothness and continuity of harmonics in the generated audio, and the filtered excitation better matches the target audio.
\end{abstract}

\section{Introduction}
\label{sec:intro}

Singing Voice Conversion (SVC) aims to convert the timbre of a song to the target singer without changing the singing content including the melody and lyrics. With the promotion of various karaoke and social Apps, SVC has been paid more and more attention to provide a better and more diverse entertainment experience. To build a good SVC system, it is expected not only to imitate the target timbre well, but also to keep the original singing content and high audio quality. Many approaches, e.g. adversarial training \cite{Nachmani2019UnsupervisedSV, 9054199} and Phonetic PosteriorGrams (PPG) \cite{saito2018non, sun2016phonetic, tian2019speaker, Chen2019SingingVC, li2021ppg}, have been proposed to extract disentangled features representing content, melody, timbre, respectively, to better convert the timbre. To produce a high-quality singing audio, the NN-based waveform generation module is usually adopted as the backend of SVC model.

Among different frameworks for audio generation, GAN-based approaches \cite{kumar2019melgan, yang2021multi, kong2020hifi, yamamoto2020parallel, zeng2021lvcnet, liu2020neural} have been widely used in VC \cite{yang2021building, wang2021noisevc, kim2021assem} and SVC \cite{guo2020phonetic, liu2021fastsvc}. Different from other architectures, such as WaveNet \cite{van2016wavenet} and WaveGlow \cite{prenger2019waveglow}, it can rapidly generate high-quality audio in parallel with a smaller-footprint model, which is helpful for the end-to-end training. However, its instability also affects the smoothness and continuity of the generated singing audio, hence causes some problems, such as pitch jitters and U/V errors in SVC. They are mainly reflected in the harmonic component which is more important in the singing voice, and affect the auditory experience of the whole song seriously. To tackle these issues, in this paper, we purpose to directly feed harmonic signals to the model to help generate the better singing voice with continuous and smooth harmonics. 

This work is composed of two aspects, the generation of harmonic signals, and the approach applying them to SVC models. Firstly, we investigate the sine excitation, which can be directly computed from F0. It is widely used in source-filter vocoders, including the neural version \cite{wang2019neural}. Then, we enhance it by filtering it with a Linear Time-Varying (LTV) filter \cite{liu2020neural}. Its coefficients are estimated by a neural network according to the input features. This signal better matches the target audio, hence improves the effect of harmonic signals. To validate the effectiveness of these signals in SVC, we conduct experiments based on the end-to-end PPG-based SVC model \cite{guo2020phonetic}. And two mainstream architectures, MelGAN \cite{kumar2019melgan} and ParallelWaveGAN \cite{yamamoto2020parallel}, are both investigated in our work, to show the universal usability of the proposed method to SVC based on adversarial audio generation.

In this paper, we will firstly introduce our approach extracting the harmonic signals in Sec.\ref{sec:harmonic_signals}, then illustrate the model architecture of the SVC model in Sec.\ref{sec:model}. Finally, a MOS test is conducted to evaluate the models in audio quality and timbre similarity. The result shows that harmonic signals enhance all SVC models significantly, especially when combining the raw sine excitation and the filtered one. In addition, the analysis of the spectrogram also reveals that the smoothness and continuity of the harmonics are improved obviously by our approach.

\section{Harmonic Signals}
\label{sec:harmonic_signals}

To provide valuable help to SVC, we first attempt to find suitable harmonic signals that well-match the target singing audio. Since the sine excitation has shown its importance in signal-based and NN-based vocoders, we mainly investigate it in our work.

\subsection{Sine Excitation}
\label{ssec:sine_excitation}

Sine excitation is a simple harmonic signal that can be directly converted from the input F0 via additive synthesis. The frame-level pitch sequence $f_0[m]$ is firstly interpolated linearly to the sample-level pitch sequence $f_0[n]$ according to the sample rate $f_s$. Then sine excitation can be generated using the below function:
\begin{equation}
p[n] = \left\{\begin{matrix}
\sum_{k=1}^{K}\sin(\phi_k[n]) & f_0[n] > 0\\ 
0 & f_0[n] = 0
\end{matrix}\right.
\end{equation}
where $K$ is the number of harmonic and $\phi_k[n]$ is the phase of the $k$-th harmonic at timestep $n$, they can be determined as follows:
\begin{equation}
K = \lfloor{\frac{f_s}{2f_0[n]}}\rfloor
\end{equation}
\begin{equation}
\phi_k[n] = \phi_k[n-1] + 2\pi k \frac{f_0[n]}{fs}
\end{equation}

\subsection{Filtered Sine Excitation}
\label{ssec:filtered_sine_excitation}

The sine excitation can describe the spectral shape of the harmonics in the singing voice accurately. However, the energy uniformly distributed on each harmonic cannot match the target audio well. The signal does not include sufficient information about the lyrics, energy, and timbre. For better matching, an LTV filter is constructed to filter the raw sine excitation based on the input features. Its coefficients are time-varied, and are estimated by a neural network according to the frame-level input sequence, which is similar to NHV \cite{liu2020neural}. This operation dynamically adjusts the amplitudes of different harmonic components, then provides a more suitable harmonic signal for the following waveform generation.

\begin{figure}[htp]
  \centering
  \includegraphics[width=7.5cm]{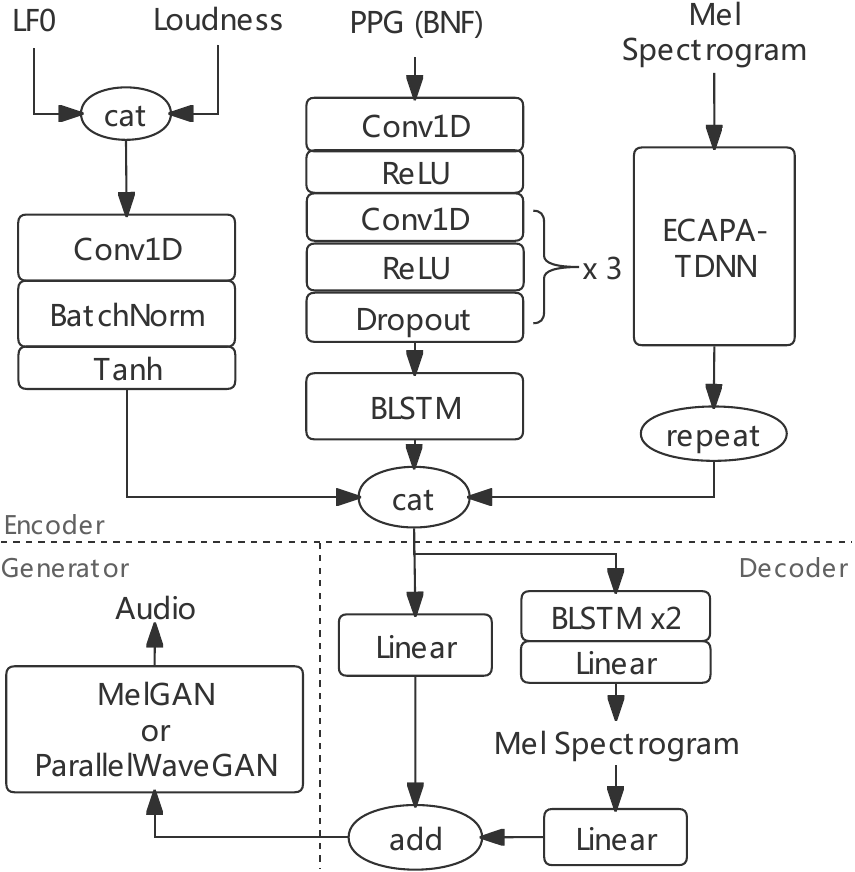}
  \caption{The architecture of our SVC model}
  \label{fig:svc}
\end{figure}

\section{Model Architecture}
\label{sec:model}

In the autoencoder-based SVC framework, feature extraction and audio reconstruction are the two main modules. The disentangled features representing content, melody, and timbre need to be extracted for precise manipulation. This work is implemented based on Phonetic PosteriorGrams (PPGs), a robust content representation, to ensure high intelligibility. In addition, for better conversion quality, an end-to-end model based on adversarial waveform generation is constructed to directly map these features to the corresponding audio.

\subsection{PPG-based Singing Voice Conversion Model}
\label{ssec:svc}

Fig.\ref{fig:svc} shows the architecture of the end-to-end SVC model. Compared with the conventional approach, which predicts acoustic features firstly, then generates audio using a vocoder trained independently, it avoids the cascade errors, and exploits richer and useful information from the input, hence generates better results \cite{guo2020phonetic}. This model is composed of three modules, an encoder, a decoder, and a waveform generation module (vocoder). They are trained jointly, and do not need pre-training.

The encoder encodes different features representing content, melody, timbre, respectively. The content representation, PPGs, is encoded through a stack of 1-D convolutional layers and a BLSTM layer, which is similar to the text encoder in Tacotron2 \cite{shen2018natural}. The LF0 and loudness non-linearly transformed are adopted as the melody representation. The ECAPA-TDNN \cite{desplanques2020ecapa} is used as the timbre encoder to extract timbre representation from the Mel-spectrogram of the input audio. Its advanced model structure and attentive statistical pooling can provide a more robust and effective speaker-dependent embedding, which has been validated in speaker verification.

Due to larger parameters and the more complicated model structure, the waveform generation module is usually unstable in end-to-end training. So we add a decoder to predict the Mel-spectrogram before it, then combine its outputs and the encoder outputs using two linear layers and an addition operation. We find that this approach can force the encoder to provide more robust acoustic-related information during the training process. In this way, the degradation of the generalization and stability due to the end-to-end training can be alleviated effectively.

\subsection{Adversarial Waveform Generation}
\label{ssec:awg}

\begin{figure}[htp]
  \centering
  \includegraphics[width=7.5cm]{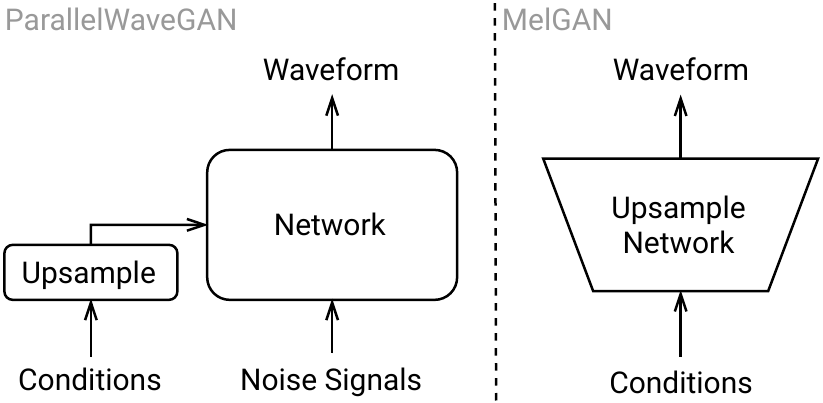}
  \caption{The architectures of MelGAN and ParallelWaveGAN}
  \label{fig:awg}
\end{figure}

As shown in Fig.\ref{fig:awg}, there are two mainstream adversarial frameworks, the MelGAN-derived and the ParallelWaveGAN-derived. MelGAN generates the waveform by upsampling the input features using a upsample network. Instead, ParallelWaveGAN (PWG) processes a Gaussian noise sequence to the target waveform with a WaveNet-like model. Most recently proposed adversarial vocoders are based on these two frameworks. So we apply our approach in both frameworks to validate its effectiveness and universal usability in adversarial waveform generation.

\begin{figure}[htp]
  \centering
  \includegraphics[width=8cm]{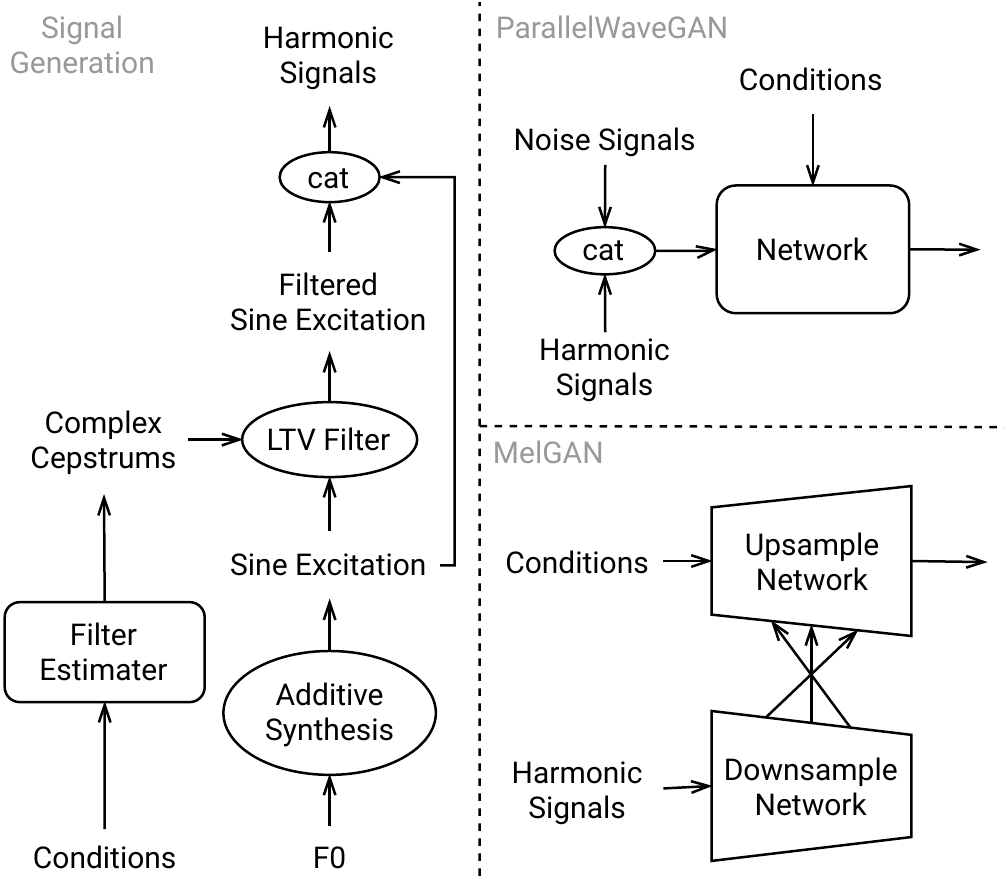}
  \caption{Applying harmonic signals to MelGAN and PWG}
  \label{fig:har_awg}
\end{figure}

The unfiltered and filtered sine excitation are both investigated in our experiments. They are inserted to MelGAN and PWG in the way shown in Fig.\ref{fig:har_awg}. Before waveform generation, the features are firstly used to estimate the LTV filter coefficients to generate harmonic signals. The F0 extracted from the source audio can be used to calculate the sine excitation. For PWG, the harmonic signals can be directly concatenated with the noise signals together as the inputs. For MelGAN, the conditions are upsampled to the hidden features with different scales by several upsampling blocks. So the harmonic signals are also downsampled to these scales via a neural network, then concatenated with them for the following operation. Due to different model structures, the filtered sine excitation is also different, which are shown in \ref{ssec:analysis}.

\begin{table*}[htb]
\centering
\caption{The MOS test results ($\pm$ indicates 95\% CI, MG: MelGAN, PWG: ParallelWaveGAN)}
\label{tab:mos}
%\resizebox{0.48\textwidth}{!}{%
\begin{tabular}{@{}C{1.2cm}|C{1.1cm}C{1.1cm}|cc|cc|cc@{}}
\toprule
\multicolumn{1}{c|}{\multirow{2}{*}{Model}} & \multicolumn{2}{c|}{Harmonic} & \multicolumn{2}{c|}{Clean} & \multicolumn{2}{c|}{Noisy} & \multicolumn{2}{c}{Overall} \\ \cline{2-9} 
\multicolumn{1}{c|}{} & Raw & Filtered & Quality & Similarity & Quality & Similarity & Quality & Similarity \\ \midrule
MG & $\times$ & $\times$ & $3.05 \pm 0.14$ & $3.17 \pm 0.14$ & $2.85 \pm 0.14$ & $3.03 \pm 0.16$ & $2.96 \pm 0.10$ & $3.10 \pm 0.11$ \\
MG & \checkmark & \checkmark & $3.45 \pm 0.14$ & $3.45 \pm 0.15$ & $3.17 \pm 0.14$ & $3.21 \pm 0.16$ & $3.31 \pm 0.10$ & $3.33 \pm 0.11$\\ \midrule
PWG    & $\times$ & $\times$ & $3.08 \pm 0.15$ & $3.22 \pm 0.16$ & $2.94 \pm 0.13$ & $3.11 \pm 0.15$ & $3.01 \pm 0.10$ & $3.17 \pm 0.11$ \\
PWG    & \checkmark & $\times$ & $3.25 \pm 0.13$ & $3.38 \pm 0.14$ & $3.07 \pm 0.14$ & $3.14 \pm 0.15$ & $3.16 \pm 0.10$ & $3.26 \pm 0.10$ \\
PWG    & \checkmark & \checkmark & $3.41 \pm 0.14$ & $3.42 \pm 0.15$ & $3.15 \pm 0.15$ & $3.19 \pm 0.15$ & $3.31 \pm 0.10$ & $3.31 \pm 0.11$ \\ \bottomrule
\end{tabular}
%}
\end{table*}

\section{EXPERIMENTS}
\label{sec:experiments}

\subsection{Experimental Protocol}
\label{ssec:setup}

Our experiments are all implemented on a clean multi-singer singing corpus. It is composed of 200 different Mandarin pop songs (total 10 hours) sung by 10 singers (5 males and 5 females, 20 songs per singer). The F0, loudness, and 80-dim Mel-spectrograms are extracted from the audio with a sample rate of 16kHz. The frameshift of these features is 10ms. PPGs are extracted from a Transformer-CTC ASR model \cite{vaswani2017attention,kim2017joint} trained with a large-scale Mandarin speech corpus (over 10,000 hours). Its 384-dim encoder output with 40ms frameshift is adopted as the PPG feature. 

In the content encoder, the first layer is a 1-D transposed convolutional layer to upsample PPGs to align other features. The output dimension of each encoder is set to 256. The structure of MelGAN is the same as \cite{guo2020phonetic}, with the upsampling factors of 8, 6, 5. The model structure of the waveform discriminator \cite{kumar2019melgan} is adopted to downsample harmonic signals to the corresponding framerates. The ParallelWaveGAN is optimized with LVCNet \cite{zeng2021lvcnet} to accelerate the training and inference speed. For the LTV filter, we adopt the model structure proposed in \cite{liu2020neural} to estimate the FIR filter coefficients to process the sine excitation.

The SVC model and the discriminator are trained adversarially with the same algorithm described in \cite{guo2020phonetic}. Besides the adversarial loss function $L_{adv}$ and the MR-STFT Loss $L_{stft}$, $L_{dec}$ is set for the decoder to calculate MAE loss between the predicted Mel-spectrogram and the target one. The final loss function is:
\begin{equation}
    L = \alpha * L_{dec} + \beta * L_{adv} + L_{stft}
\end{equation}
which $\alpha$ and $\beta$ are set to 200 and 4 respectively. All models are trained for 400,000 iterations with a batch size of 16x4, i.e. 16 audio segments with the length of 4 seconds.\footnote{The details of LVCNet can be found at \url{https://github.com/zceng/LVCNet}.} \footnote{The training details can be found at \url{https://github.com/hhguo/EA-SVC}}

\subsection{Subjective Evaluation}
\label{ssec:mos}

The mean opinion score (MOS) test is conducted to evaluate the proposed method in two aspects: sound quality and singer similarity. Two test sets are used, a clean set with 30 high-quality clean singing segments, and a noisy set with 30 singing segments recorded with worse quality. The timbre in each segment is converted to a random singer contained in the training set. Finally, 24 listeners are involved to rate the 60 test cases. Each of them only rates 20 cases to ensure an accurate result \footnote{Samples are available at \url{https://hhguo.github.io/DemoHarSVC}}.

The final results are shown in Table.\ref{tab:mos}. For the same model structure, the models with harmonic signals achieve the higher score on all test sets, especially for the clean set, which shows the best conversion performance. For the noisy set, although the extraction of PPG and F0 are easily contaminated by the noise, it still achieves significant improvement, which validates the effectiveness and robustness of our method. Moreover, compared with the similarity, the improvement is more obvious in sound quality, which gets an increase of more than 0.3 on the whole test set. It strongly verifies that it is important to present good harmonics in SVC.

To investigate the effect of the filtered sine excitation, we also evaluate the PWG with or without it. The MOS result shows that the model with only the raw sine excitation can already outperform the baseline PWG. But the filtered sine excitation can further improve the model significantly, which shows its importance and improvement to SVC.

\begin{figure}[htp]
  \centering
  \includegraphics[width=8cm]{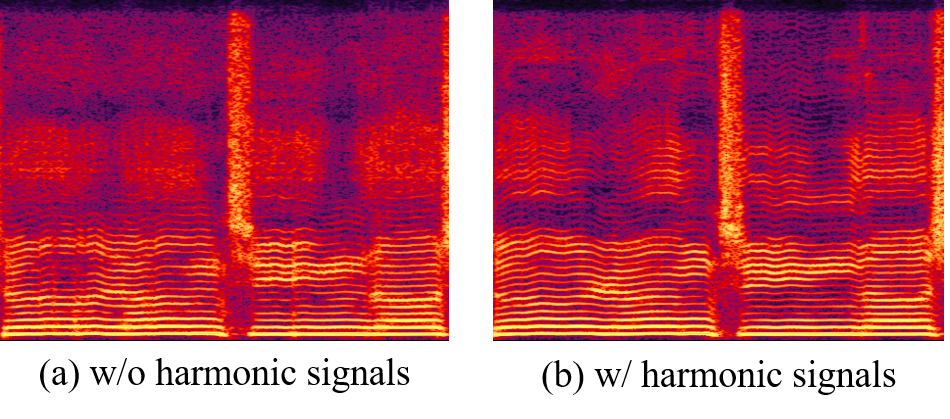}
  \caption{The spectrograms generated by the MelGAN models w/ or w/o harmonic signals}
  \label{fig:spectrograms}
\end{figure}

\subsection{Analysis}
\label{ssec:analysis}

Fig.\ref{fig:spectrograms} shows the spectrograms synthesized by MelGAN with or without harmonic signals. The problems in (a), including pitch jitter and unsmoothed harmonics, are all solved in (b). Besides, different from the muffle spectrogram in the middle frequency in (a), (b) shows more clear harmonics, which makes the converted singing audio more smooth and clean.

\begin{figure}[htp]
  \centering
  \includegraphics[width=8.5cm]{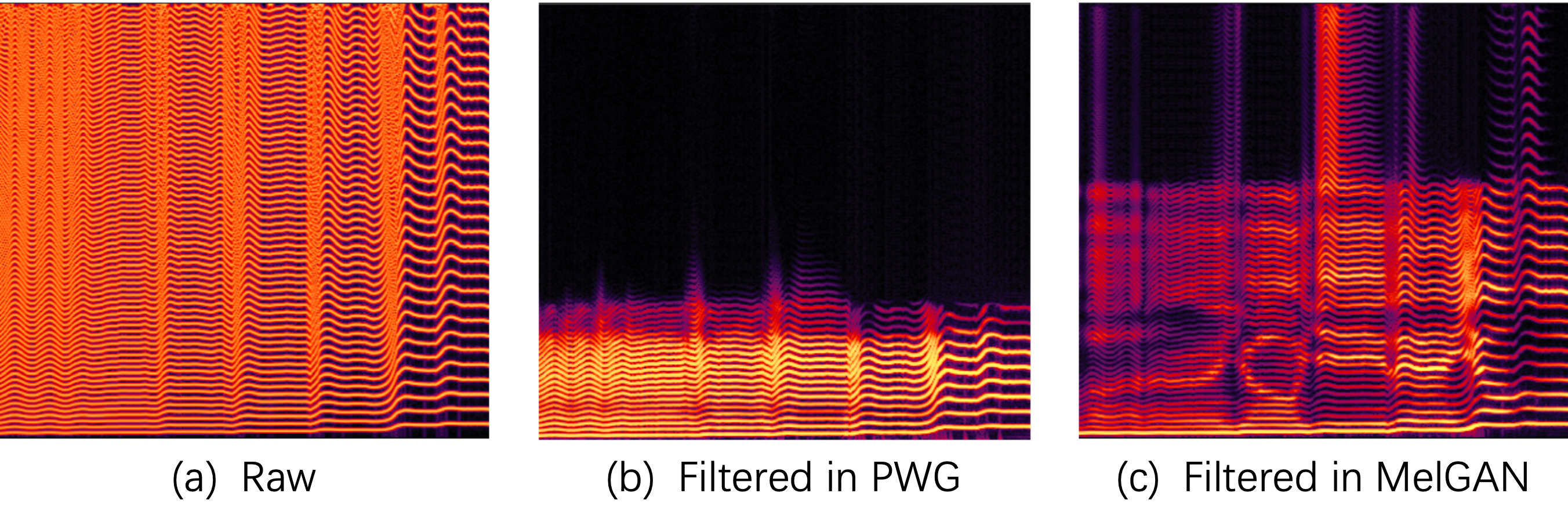}
  \caption{The spectrograms of different sine excitations}
  \label{fig:filter_sin}
\end{figure}

We also compare different harmonic signals in MelGAN and PWG. As shown in Fig.\ref{fig:filter_sin}, (a) is the spectrogram of the raw sine excitation. It only has harmonics with evenly distributed energy. After filtered by the LTV filter in PWG, the high frequency is removed dynamically, whose range is different at each frame. In the reserved part, the energy is distributed differently to each curve, which better matches the target audio. This phenomenon becomes more eminent in MelGAN. In (c), the reserved frequency has a wider range and higher variety. The vowel is synthesized well that can be recognized clearly. Its timbre also becomes more distinguishable to imitate the target one.

\section{CONCLUSION}
\label{sec:conclusion}

This paper purposes to utilize harmonic signals to improve singing voice conversion based on adversarial waveform generation. We investigate two signals, the raw sine excitation and the sine excitation filtered by an estimated LTV filter, and apply them to two mainstream adversarial waveform generation models, MelGAN and ParallelWaveGAN. The experimental results show our method significantly improves the sound quality and timbre similarity on the clean and noisy test sets. And the filtered sine excitation is also validated as a powerful harmonic signal to SVC. The case analysis shows that the method improves the smoothness, continuity, and clarity of harmonic components. And the filtered excitation better matches the target audio.

%\vfill
%\pagebreak
\vfill\pagebreak

\small
\bibliographystyle{IEEEbib}
\bibliography{refs}

\end{document}